\documentclass{article}
\usepackage{epsfig}
\usepackage{amsmath}

\input{tcilatex}

\begin{document}

\title{Quantum games with correlated noise}
\author{Ahmad Nawaz\thanks{%
ahmad@ele.qau.edu.pk} and A. H. Toor\thanks{%
ahtoor@qau.edu.pk} \\
Department of Physics, Quaid-i-Azam University, \ \ \ \ \ \\
Islamabad 45320, Pakistan.}
\maketitle

\begin{abstract}
We analyze quantum game with correlated noise through generalized
quantization scheme. Four different combinations on the basis of
entanglement of initial quantum state and the measurement basis are
analyzed. It is shown that the advantage that a quantum player can get by
exploiting quantum strategies is only valid when both the initial quantum
state and the measurement basis are in entangled form. Furthermore, it is
shown that for maximum correlation the effects of decoherence diminish and
it behaves as a noiseless game.
\end{abstract}

\section{Introduction}

It requires exchange of qubits between arbiter and players to play quantum
games. The\ transmission of qubit \ through a channel is generally prone to
decoherence due to its interaction with the environment. In the game
theoretic sense this situation can be imagined as if a demon is present
between the arbiter and the players who corrupts the qubits. The players are
not necessarily aware of the actions of the demon \cite{lee}. This type of
protocol was first applied to quantum games to show that above a certain
level of decoherence the quantum player has no advantage over a classical
player \cite{johnson}\emph{. }Later quantum version of Prisoners' Dilemma
was analyzed in presence of decoherence to prove that Nash equilibrium is
not affected by decoherence \cite{chen}. Recently, Flitney and Abbott \cite%
{flitney} showed for the quantum games based on dephasing quantum channel
that the advantage that a quantum player enjoys over a classical player
diminishes as decoherence increases and vanishes for the maximum decoherence.

In this paper we analyze the quantum games based on quantum correlated
dephasing channel in the context of our generalized quantization scheme for
non-zero sum games \cite{nawaz}. We identified four different combinations
on the basis of initial state entanglement parameter, $\gamma ,$\ and the
measurement parameter, $\delta ,$\ \ for some quantum games.\emph{\ }It is
shown that for $\gamma =\delta =0$ the game reduce to the classical and\emph{%
\ }become independent of decoherence and memory effects. For the case when $%
\gamma \neq 0,\delta =0$ the scheme reduces to Marinatto and Weber
quantization scheme \cite{marinatto}. It is interesting to note that though
the initial state is entangled, quantum player has no advantage over the
classical player.\emph{\ }Same happens for the case of $\gamma =0,\delta
\neq 0$. However, for the case when $\gamma =\delta =\frac{\pi }{2}$ the
scheme transforms to the Eisert's quantization scheme \cite{eisert} and
quantum player always remains better off against a player restricted to
classical strategies. Furthermore, in the limit of maximum correlation the
effect of decoherence vanishes and the quantum game behaves as a noiseless
game.

The paper is organized as follows: Sec. \ref{decoherence}\ deals with
quantization of quantum games in presence of correlated noise and a brief
introduction to some classical games of interest is given in the Appendix A.

\section{\label{decoherence}Quantization in presence of correlated noise}

Decoherence is a non-unitary dynamics that results due to the coupling of
principal system with the environment. One of the important type of
decoherence is phase damping or dephasing. It is uniquely quantum mechanical
and describes the loss of quantum information without loss of energy. The
energy eigenstate of the system do not change\ as a function of time during
this process but the system accumulates a phase proportional to the
eigenvalue. With the passage of time the relative phase between the energy
eingenstates may lost.

In pure dephasing process a qubit transforms as 
\begin{equation}
c\left| 0\right\rangle +b\left| 1\right\rangle \rightarrow c\left|
0\right\rangle +be^{i\phi }\left| 1\right\rangle  \label{initial state}
\end{equation}%
where $\phi $\ is the phase kick. If this phase kick, $\phi $ is assumed to
be a random variable with Gaussian distribution of mean zero and variance $%
2\lambda $ then the density matrix of system after averaging over all the
values of $\phi $ is \cite{chuang}%
\begin{equation}
\left[ 
\begin{array}{cc}
\left| a\right| ^{2} & ab^{\ast } \\ 
a^{\ast }b & \left| b\right| ^{2}%
\end{array}%
\right] \rightarrow \left[ 
\begin{array}{cc}
\left| a\right| ^{2} & ab^{\ast }e^{-\lambda } \\ 
a^{\ast }be^{-\lambda } & \left| b\right| ^{2}%
\end{array}%
\right]  \label{dephasing}
\end{equation}%
It is evident from the above equation that in this process the phase kicks
cause the\emph{\ }off-diagonal elements of the density matrix to decay
exponentially to zero with time. In the operator sum representation\ the
dephasing process can be expressed as \cite{kraus,chuang} 
\begin{equation}
\rho _{f}=\overset{1}{\underset{i=0}{\tsum }}A_{i\text{ }}\rho _{in}\text{ }%
A_{i}^{\dagger }  \label{kraus-1}
\end{equation}%
where\emph{\ } 
\begin{eqnarray}
A_{0} &=&\sqrt{1-\frac{p}{2}}I  \notag \\
A_{1} &=&\sqrt{\frac{p}{2}}\sigma _{z}  \label{kraus}
\end{eqnarray}%
are the Kraus operators$,\ I$ is the identity operator and $\sigma _{z}$\ is
the Pauli matrix. Recognizing\emph{\ }$1-p=e^{-\lambda },$ let $N$\ qubits
are allowed to pass through such a channel then the Eq. \ref{kraus-1}\
becomes 
\begin{equation}
\rho _{f}=\overset{}{\underset{k_{1},\text{ ...,\ }k_{n}=0}{\tsum }}\left(
A_{k_{n}}\otimes ......A_{k_{1}}\right) \rho _{in\text{ }}\left(
A_{k_{1}}^{\dagger }\otimes ......A_{k_{n}}^{\dagger }\right) .
\label{kraus-operators}
\end{equation}%
Now if noise is correlated with memory of degree $\mu ,$ the Kraus operator
for two qubit system becomes \cite{palma}

\begin{equation}
A_{i,j}=\sqrt{p_{i}\left[ \left( 1-\mu \right) p_{j}+\mu \delta _{ij}\right] 
}\sigma _{i}\otimes \sigma _{j}.  \label{kraus-memory}
\end{equation}%
where $i,j=0$ and $z$ with $\sigma _{o}=I.$ Physically, this expression
means that with the\emph{\ }probability $1-\mu $ the noise is uncorrelated
and can be completely specified by the Kraus operators $A_{i,j}^{u}=\sqrt{%
p_{i}p_{j}}\sigma _{i}\otimes \sigma _{j}$ whereas with probability $\mu $
the noise is correlated and is specified by Kraus operators of the form $%
A_{ii}^{c}=\sqrt{p_{i}}\sigma _{i}\otimes \sigma _{i}.$

The protocol for quantum games in presence of decoherence is developed in
the Ref. \cite{flitney}. An initial entangled state is prepared by the
arbiter and passed on to the players through a dephasing quantum channel. On
receiving the quantum state players apply their local operators (strategies)
and return it back to arbiter through dephasing quantum channel. Then
arbiter performs the measurement and announces their payoffs.

Let the game starts with the initial quantum state: 
\begin{equation}
\left| \psi _{in}\right\rangle =\cos \frac{\gamma }{2}\left| 00\right\rangle
+i\sin \frac{\gamma }{2}\left| 11\right\rangle .  \label{initialstate}
\end{equation}%
The strategies of the players in the generalized quantization scheme\emph{\ }%
is represented by the unitary operator $U_{i}$\ of the form \cite{nawaz} 
\begin{equation}
U_{i}=\cos \frac{\theta _{i}}{2}R_{i}+\sin \frac{\theta _{i}}{2}P_{i},\text{
\ \ \ }  \label{combination}
\end{equation}%
where $i=1$\ or $2$\ and $R_{i}$, $P_{i}$\emph{\ }are the unitary operators
defined as:

\begin{align}
R_{i}\left| 0\right\rangle & =e^{i\alpha _{i}}\left| 0\right\rangle ,\text{
\ \ }R_{i}\left| 1\right\rangle =e^{-i\alpha _{i}}\left| 1\right\rangle , 
\notag \\
P_{i}\left| 0\right\rangle & =e^{i\left( \frac{\pi }{2}-\beta _{i}\right)
}\left| 1\right\rangle ,\text{ \ \ \ \ \ }P_{i}\left| 1\right\rangle
=e^{i\left( \frac{\pi }{2}+\beta _{i}\right) }\left| 0\right\rangle ,
\label{operators}
\end{align}%
where $0\leq \theta \leq \pi ,-\pi \leq \alpha ,\beta \leq \pi .$\emph{\ }%
Here we extended our earlier generalized quantization scheme to three
strategy set of parameters in accordance with Ref. \cite{flitney}\emph{.\ }%
After the application of these strategies, the initial state given by the
Eq. (\ref{initialstate}) transforms to 
\begin{equation}
\rho _{f}=(U_{1}\otimes U_{2})\rho _{in}(U_{1}\otimes U_{2})^{\dagger },
\label{final state}
\end{equation}%
where $\rho _{in}=\left| \psi _{in}\right\rangle \left\langle \psi
_{in}\right| $ is the density matrix for the quantum state. The operators
used by the arbiter to determine the payoff for Alice and Bob are%
\begin{equation}
P=\$_{00}P_{00}+\$_{01}P_{01}+\$_{10}P_{10}+\$_{11}P_{11},
\label{payoff operator}
\end{equation}%
where 
\begin{subequations}
\label{oper a}
\begin{align}
P_{00}& =\left| \psi _{00}\right\rangle \left\langle \psi _{00}\right| \text{%
, \ }\left| \psi _{00}\right\rangle =\cos \left( \delta /2\right) \left|
00\right\rangle +i\sin \left( \delta /2\right) \left| 11\right\rangle ,
\label{oper 1} \\
P_{11}& =\left| \psi _{11}\right\rangle \left\langle \psi _{11}\right| \text{%
, \ }\left| \psi _{11}\right\rangle =\cos \left( \delta /2\right) \left|
11\right\rangle +i\sin \left( \delta /2\right) \left| 00\right\rangle ,
\label{oper 2} \\
P_{10}& =\left| \psi _{10}\right\rangle \left\langle \psi _{10}\right| \text{%
, \ }\left| \psi _{10}\right\rangle =\cos \left( \delta /2\right) \left|
10\right\rangle -i\sin \left( \delta /2\right) \left| 01\right\rangle ,
\label{oper 3} \\
P_{01}& =\left| \psi _{01}\right\rangle \left\langle \psi _{01}\right| \text{%
, \ }\left| \psi _{01}\right\rangle =\cos \left( \delta /2\right) \left|
01\right\rangle -i\sin \left( \delta /2\right) \left| 10\right\rangle ,
\label{oper 4}
\end{align}%
with\emph{\ }$\delta \in \left[ 0,\frac{\pi }{2}\right] $ and $\$_{ij}$ are
the elements of payoff matrix in the $ith$ row and $jth$ column (given in
Appendix A for different games). Above payoff operators reduce to that of
Eisert's scheme for $\delta $ equal to $\gamma ,$ which represents the
entanglement of the initial state \cite{eisert}. And for $\delta =0$ above
operators transform into that of Marinatto and Weber's scheme \cite%
{marinatto}. In our extended\ generalized quantization to three set of
parameters scheme, payoffs for the players are: 
\end{subequations}
\begin{eqnarray}
\$^{A}(\theta _{i},\alpha _{i},\beta _{i}) &=&\text{Tr}(P_{A}\rho _{f})\text{%
,}  \notag \\
\$^{B}(\theta _{i},\alpha _{i},\beta _{i}) &=&\text{Tr}(P_{B}\rho _{f}),
\label{payoff formula}
\end{eqnarray}%
\emph{\ } where Tr represents the trace of a\emph{\ }matrix. Using Eqs. (\ref%
{kraus-memory}) (\ref{initialstate}), (\ref{payoff operator}), and (\ref%
{payoff formula}), the payoffs come out to be 
\begin{eqnarray}
\$(\theta _{i},\alpha _{i},\beta _{i}) &=&c_{1}c_{2}\left[ \eta \$_{00}+\chi
\$_{11}+\left( \$_{00}-\$_{11}\right) \mu _{p}^{(1)}\mu _{p}^{(2)}\xi \cos
2(\alpha _{1}+\alpha _{2})\right]   \notag \\
&&+s_{1}s_{2}\left[ \eta \$_{11}+\chi \$_{00}-\left( \$_{00}-\$_{11}\right)
\mu _{p}^{(1)}\mu _{p}^{(2)}\xi \cos 2(\beta _{1}+\beta _{2})\right]   \notag
\\
&&+c_{1}s_{2}\left[ \eta \$_{01}+\chi \$_{10}+\left( \$_{01}-\$_{10}\right)
\mu _{p}^{(1)}\mu _{p}^{(2)}\xi \cos 2(\alpha _{1}-\beta _{2})\right]  
\notag \\
&&+c_{2}s_{1}\left[ \eta \$_{10}+\chi \$_{01}-\left( \$_{01}-\$_{10}\right)
\mu _{p}^{(1)}\mu _{p}^{(2)}\xi \cos 2(\alpha _{2}-\beta _{1})\right]  
\notag \\
&&+\frac{\mu _{p}^{(2)}\left( \$_{00}-\$_{11}\right) }{4}\sin \theta
_{1}\sin \theta _{2}\sin \delta \sin \left( \alpha _{1}+\alpha _{2}+\beta
_{1}+\beta _{2}\right)   \notag \\
&&+\frac{\mu _{p}^{(2)}\left( \NEG{\$}_{10}-\$_{01}\right) }{4}\sin \theta
_{1}\sin \theta _{2}\sin \delta \sin \left( \alpha _{1}-\alpha _{2}+\beta
_{1}-\beta _{2}\right)   \notag \\
&&+\frac{\mu _{p}^{(1)}\left( -\$_{00}-\$_{11}+\$_{01}+\$_{10}\right) }{4}%
\sin \theta _{1}\sin \theta _{2}\sin \gamma \sin \left( \alpha _{1}+\alpha
_{2}-\beta _{1}-\beta _{2}\right)   \notag \\
&&  \label{payoff}
\end{eqnarray}%
where 
\begin{subequations}
\label{abrs}
\begin{eqnarray*}
\eta  &=&\cos ^{2}\left( \delta /2\right) \cos ^{2}\left( \gamma /2\right)
+\sin ^{2}\left( \delta /2\right) \sin ^{2}\left( \gamma /2\right) , \\
\chi  &=&\cos ^{2}\left( \delta /2\right) \sin ^{2}\frac{\gamma }{2}+\sin
^{2}\left( \delta /2\right) \cos ^{2}\left( \gamma /2\right) , \\
\xi  &=&1/2\left( \sin \delta \sin \gamma \right) , \\
c_{i} &=&\cos ^{2}\frac{\theta _{i}}{2}, \\
s_{i} &=&\sin ^{2}\frac{\theta _{i}}{2} \\
\mu _{p}^{(i)} &=&\left( 1-\mu _{i}\right) \left( 1-p_{i}\right) ^{2}+\mu
_{i}.
\end{eqnarray*}%
The payoff for the two players can be found by putting the appropriate
values for $\$_{ij}$ (elements of the payoff matrix for the corresponding
game) in the Eq. \ref{payoff}. These payoffs become the classical payoffs
for $\delta =\gamma =0$ and for $\delta =\gamma =\frac{\pi }{2}$and $\mu =0$
these payoffs transform to the results given in the\emph{\ }Ref \cite%
{flitney}. It is known that decoherence has no effect on the Nash
equilibrium of the game but it causes a reduction in the payoffs \cite%
{chen,flitney}. In our case it is interesting to note that this reduction of
the payoffs depends on the degree of memory $\mu .$ As $\mu $ increases from
zero to one the effect of noise reduces until finally for $\mu =1$ the
payoffs become as that for noiseless game irrespective of any value of $p_{i}
$. It is further to be noted that in comparison to memoryless case \cite%
{flitney} the quantum phases $\alpha _{i},\beta _{i}$\ do not vanish even
for maximum value of decoherence\emph{,} i.e.\emph{,.} for $p_{1}=p_{2}=1$.

To see further the effects of memory in quantum games we consider a
situation in which Alice is restricted to play classical strategies, i.e., $%
\alpha _{1}=\beta _{1}=0,$ whereas Bob is capable of playing the quantum
strategies as well. Under these circumstances following four cases for the
different combinations of $\delta $ and $\gamma $ are worth noting:

\textbf{Case (i)} When $\delta =\gamma =0$ then it is clear from the\emph{\ }%
Eq. (\ref{payoff}) payoffs \ are the same as in the case of\emph{\ }%
classical game \cite{eisert1}. These payoffs, as expected, are independent
of the dephasing probabilities $p_{i}$, the quantum strategies $\alpha
_{2},\beta _{2}$ and the memory.

\textbf{Case (ii)} When $\delta =0,\gamma \neq 0$ then $\eta =\cos ^{2}\frac{%
\gamma }{2},\chi =\sin ^{2}\frac{\gamma }{2},$and $\xi =0.$Using payoff
matrix for the game of Prisoners Dilemma, given in Appendix A, and the Eq.%
\emph{\ }(\ref{payoff}) the payoffs for the two players are: 
\end{subequations}
\begin{eqnarray}
\$^{A}(\theta _{1},\theta _{2},\alpha _{2},\beta _{2}) &=&c_{1}c_{2}\left(
3-2\sin ^{2}\frac{\gamma }{2}\right) +s_{1}s_{2}\left( 1+2\sin ^{2}\frac{%
\gamma }{2}\right)  \notag \\
&&+5c_{1}s_{2}\sin ^{2}\frac{\gamma }{2}+5c_{2}s_{1}\left( 1-\sin ^{2}\frac{%
\gamma }{2}\right)  \notag \\
&&+\frac{\mu _{p}^{(1)}}{4}\sin \theta _{1}\sin \theta _{2}\sin \gamma \sin
\left( \alpha _{2}-\beta _{2}\right)  \notag \\
\$^{B}(\theta _{1},\theta _{2},\alpha _{2},\beta _{2}) &=&c_{1}c_{2}\left(
3-2\sin ^{2}\frac{\gamma }{2}\right) +s_{1}s_{2}\left( 1+2\sin ^{2}\frac{%
\gamma }{2}\right)  \notag \\
&&+5c_{1}s_{2}\left( 1-\sin ^{2}\frac{\gamma }{2}\right) +5c_{2}s_{1}\sin
^{2}\frac{\gamma }{2}  \notag \\
&&+\frac{\mu _{p}^{(1)}}{4}\sin \theta _{1}\sin \theta _{2}\sin \gamma \sin
\left( \alpha _{2}-\beta _{2}\right)
\end{eqnarray}%
\ In this case the optimal strategy for the quantum player, Bob, is $\alpha
_{2}-\beta _{2}=\frac{\pi }{2}.$ Though his choice for $\theta _{2}$ depends
on Alice's choice for\emph{\ }$\theta _{1},$but he can play $\theta _{2}=%
\frac{\pi }{2},$ without being bothered about Alice's choice as rational
reasoning leads Alice to play\emph{\ }$\theta _{1}=\frac{\pi }{2}$. Under
these choices of moves the payoffs for the two players are equal: 
\begin{eqnarray}
\$^{A}(\frac{\pi }{2},\frac{\pi }{2},\alpha _{2}-\beta _{2} &=&\frac{\pi }{2}%
)=\$^{B}(\frac{\pi }{2},\frac{\pi }{2},\alpha _{2}-\beta _{2}=\frac{\pi }{2})
\notag \\
&=&\frac{9}{4}+\frac{\mu _{p}^{(1)}}{4}\sin \gamma .
\end{eqnarray}%
It is evident that the quantum player has no advantage over the classical
player. Similarly for the Chicken game the payoffs for the two players are: 
\begin{eqnarray}
\$^{A}(\theta _{1},\theta _{2},\alpha _{2},\beta _{2}) &=&c_{1}c_{2}\left(
3-3\sin ^{2}\frac{\gamma }{2}\right) +s_{1}s_{2}\left( 3\sin ^{2}\frac{%
\gamma }{2}\right)  \notag \\
&&+c_{1}s_{2}\left( 3\sin ^{2}\frac{\gamma }{2}+1\right) +c_{2}s_{1}\left(
4-3\sin ^{2}\frac{\gamma }{2}\right)  \notag \\
&&+\frac{\mu _{p}^{(1)}}{2}\sin \theta _{1}\sin \theta _{2}\sin \gamma \sin
\left( \alpha _{2}-\beta _{2}\right) \\
\$^{B}(\theta _{1},\theta _{2},\alpha _{2},\beta _{2}) &=&c_{1}c_{2}\left(
3-3\sin ^{2}\frac{\gamma }{2}\right) +s_{1}s_{2}\left( 3\sin ^{2}\frac{%
\gamma }{2}\right)  \notag \\
&&+c_{1}s_{2}\left( 4-3\sin ^{2}\frac{\gamma }{2}\right) +c_{2}s_{1}\left(
1+3\sin ^{2}\frac{\gamma }{2}\right)  \notag \\
&&+\frac{\mu _{p}^{(1)}}{2}\sin \theta _{1}\sin \theta _{2}\sin \gamma \sin
\left( \alpha _{2}-\beta _{2}\right)
\end{eqnarray}%
and it can be shown using the same argument as for the game of Prisoner
Dilemma that the quantum player does not have any advantage over classical
player in the Chicken game as well.

For the case of the quantum Battle of Sexes the payoffs become 
\begin{eqnarray}
\$^{A}(\theta _{1},\theta _{2},\alpha _{2},\beta _{2}) &=&c_{1}c_{2}\left(
2-\sin ^{2}\frac{\gamma }{2}\right) +s_{1}s_{2}\left( 1+\sin ^{2}\frac{%
\gamma }{2}\right)  \notag \\
&&-\frac{3\mu _{p}^{(1)}}{4}\sin \theta _{1}\sin \theta _{2}\sin \gamma \sin
\left( \alpha _{2}-\beta _{2}\right)  \notag \\
\$^{B}(\theta _{1},\theta _{2},\alpha _{2},\beta _{2}) &=&c_{1}c_{2}\left(
1+\sin ^{2}\frac{\gamma }{2}\right) +s_{1}s_{2}\left( 2-\sin ^{2}\frac{%
\gamma }{2}\right)  \notag \\
&&-\frac{3\mu _{p}^{(1)}}{4}\sin \theta _{1}\sin \theta _{2}\sin \gamma \sin
\left( \alpha _{2}-\beta _{2}\right) .
\end{eqnarray}%
Here the optimal strategy for Bob is $\alpha _{2}-\beta _{2}=-\frac{\pi }{2}$
and $\theta _{2}=\frac{\pi }{2}$, keeping in view that the best strategy for
Alice is $\theta _{1}=\frac{\pi }{2}.$ The corresponding payoffs of the
players are again equal for these choices, i.e., 
\begin{eqnarray}
\$^{A}(\frac{\pi }{2},\frac{\pi }{2},\alpha _{2}-\beta _{2} &=&-\frac{\pi }{2%
})=\$^{B}(\frac{\pi }{2},\frac{\pi }{2},\alpha _{2}-\beta _{2}=-\frac{\pi }{2%
})  \notag \\
&=&\frac{3}{4}+\frac{3}{4}\mu _{p}^{(1)}\sin \gamma
\end{eqnarray}%
It is clear that for the case $\delta =0,\gamma \neq 0$ the quantum player
has no advantage over the classical player for three games considered above%
\emph{. }It is interesting\emph{\ }because\emph{\ }the game starts from an
entangled state and the payoffs are also the functions of the quantum
phases, $\alpha _{i},\beta _{i},$ dephasing probability, $p_{1}$ and the
degree of memory, $\mu _{1}$, of the quantum channel between Bob and arbiter.

\textbf{Case (iii)} When $\delta \neq 0,\gamma =0$\ then using Eq. \ref%
{payoff}, the payoffs for the two players in games of Prisoners Dilemma,
Chicken and Battle of sexes are 
\begin{eqnarray}
\$^{A}(\theta _{1},\theta _{2},\alpha _{2},\beta _{2}) &=&c_{1}c_{2}\left(
3-2\sin ^{2}\frac{\delta }{2}\right) +s_{1}s_{2}\left( 1+2\sin ^{2}\frac{%
\delta }{2}\right)  \notag \\
&&+\frac{7\mu _{p}^{(2)}}{4}\sin \theta _{1}\sin \theta _{2}\sin \delta \sin
\left( \alpha _{2}+\beta _{2}\right)  \notag \\
\$^{B}(\theta _{1},\theta _{2},\alpha _{2},\beta _{2}) &=&c_{1}c_{2}\left(
1+\sin ^{2}\frac{\delta }{2}\right) +s_{1}s_{2}\left( 2-\sin ^{2}\frac{%
\delta }{2}\right)  \notag \\
&&-\frac{3\mu _{p}^{(2)}}{4}\sin \theta _{1}\sin \theta _{2}\sin \delta \sin
\left( \alpha _{2}+\beta _{2}\right) ,
\end{eqnarray}%
\begin{eqnarray}
\$^{A}(\theta _{1},\theta _{2}) &=&c_{1}c_{2}\left( 3-3\sin ^{2}\frac{\delta 
}{2}\right) +s_{1}s_{2}\left( 3\sin ^{2}\frac{\delta }{2}\right)
+c_{1}s_{2}\left( 1+3\sin ^{2}\frac{\delta }{2}\right)  \notag \\
&&+c_{2}s_{1}\left( 4-3\sin ^{2}\frac{\delta }{2}\right)  \notag \\
\$^{B}(\theta _{1},\theta _{2},\alpha _{2},\beta _{2}) &=&c_{1}c_{2}\left(
3-3\sin ^{2}\frac{\delta }{2}\right) +s_{1}s_{2}\left( 3\sin ^{2}\frac{%
\delta }{2}\right) +c_{1}s_{2}\left( 4-3\sin ^{2}\frac{\delta }{2}\right) 
\notag \\
&&+c_{2}s_{1}\left( 1+3\sin ^{2}\frac{\delta }{2}\right) +\frac{3\mu
_{p}^{(2)}}{2}\sin \theta _{1}\sin \theta _{2}\sin \delta \sin \left( \alpha
_{2}+\beta _{2}\right) ,  \notag \\
&&
\end{eqnarray}%
\emph{\ }%
\begin{eqnarray}
\$^{A}(\theta _{1},\theta _{2},\alpha _{2},\beta _{2}) &=&c_{1}c_{2}\left(
2-\sin ^{2}\frac{\delta }{2}\right) +s_{1}s_{2}\left( 1+\sin ^{2}\frac{%
\delta }{2}\right)  \notag \\
&&+\frac{3\mu _{p}^{(2)}}{4}\sin \theta _{1}\sin \theta _{2}\sin \delta \sin
\left( \alpha _{2}+\beta _{2}\right)  \notag \\
\$^{B}(\theta _{1},\theta _{2},\alpha _{2},\beta _{2}) &=&c_{1}c_{2}\left(
1+\sin ^{2}\frac{\delta }{2}\right) +s_{1}s_{2}\left( 2-\sin ^{2}\frac{%
\delta }{2}\right)  \notag \\
&&-\frac{3\mu _{p}^{(2)}}{4}\sin \theta _{1}\sin \theta _{2}\sin \delta \sin
\left( \alpha _{2}+\beta _{2}\right) ,
\end{eqnarray}%
respectively.\emph{\ }It is evident from the above expressions for the
payoffs that the optimal strategy for Bob, the quantum player, is $\alpha
_{2}+\beta _{2}=-\frac{\pi }{2}$ with,\ $\theta _{2}=\frac{\pi }{2},$for
Prisoners Dilemma and Battle of sexes. But corresponding payoff for Alice is
less. However, she can overcome this by playing $\theta _{1}=0$ or $\pi ,$\
so that the payoffs for both the players become independent of the quantum
phases $\alpha _{2},\beta _{2}$. So there remain no option for the quantum
player to enhance his payoff by exploiting the quantum move. However in the
case of chicken game the quantum player can enhance his payoff without
effecting the payoff of classical player. But again the classical player has
the ability to prevent quantum strategies by playing $\theta _{1}=0$ or $\pi
.$ So there remains no advantage for playing quantum strategies. It is also
interesting to note that though by playing this move Alice could force the
payoffs of the two players to be independent of dephasing factor $p_{2}$\
and the degree of memory $\mu _{2}$, however, the game remains different
from its classical counterpart.

\textbf{Case (iv)} When $\delta =\gamma =\frac{\pi }{2},$then Eq.\ref{payoff}
with $\mu _{1}=\mu _{2}=0,$ gives the same results as mentioned in Ref. \cite%
{flitney} and the quantum player is better off for $p<1$. However, when
decoherence increases this advantage diminishes and vanishes for maximum
decoherence,\emph{\ }i.e.,\emph{\ }$p=1$. But in our case when $\mu \neq 0$,
the quantum player is always better off even for maximum noise, i.e., $p=1,$
which was not possible in memoryless case$.$ Furthermore it is worth noting
that as the degree of memory increases from $0$ to $1$ the effect of noise
on the payoffs starts decreasing and for $\mu =1$ it behaves like a
noiseless game.

In the case of Prisoners Dilemma, the optimal strategy for Bob is to play $%
\alpha _{2}=\frac{\pi }{2}$ and $\beta _{2}=0.$ His choice for, $\theta
_{2}, $ is $\frac{\pi }{2}$, independent of Alice's move. The payoffs for
Alice and Bob as a function of decoherence probability $p_{1}=p_{2}=p$ at $%
\mu =\frac{1}{2},$ is%
\begin{eqnarray}
\$^{A}(\theta _{1},\theta _{2},\alpha _{2},\beta _{2}) &=&c_{1}c_{2}\left[
2+\mu _{p}^{2}\cos 2\alpha _{2}\right] +s_{1}s_{2}\left[ 2-\mu _{p}^{2}\cos
2\beta _{2}\right]  \notag \\
&&+\frac{5}{2}c_{1}s_{2}\left[ 1-\mu _{p}^{2}\cos 2\beta _{2})\right] +\frac{%
5}{2}c_{2}s_{1}\left[ 1+\mu _{p}^{2}\cos 2\alpha _{2}\right]  \notag \\
&&+\frac{\mu _{p}}{4}\sin \theta _{1}\sin \theta _{2}\sin \left( \alpha
_{2}-\beta _{2}\right) -\frac{3\mu _{p}}{4}\sin \theta _{1}\sin \theta
_{2}\sin \left( \alpha _{2}+\beta _{2}\right)  \notag \\
&&
\end{eqnarray}%
\begin{eqnarray}
\$^{B}(\theta _{1},\theta _{2},\alpha _{2},\beta _{2}) &=&c_{1}c_{2}\left[
2+\mu _{p}^{2}\cos 2\alpha _{2}\right] +s_{1}s_{2}\left[ 2-\mu _{p}^{2}\cos
2\beta _{2}\right]  \notag \\
&&+\frac{5}{2}c_{1}s_{2}\left[ 1+\mu _{p}^{2}\cos 2\beta _{2}\right] +\frac{5%
}{2}c_{2}s_{1}\left[ 1-\mu _{p}^{2}\cos 2\alpha _{2}\right]  \notag \\
&&+\frac{7\mu _{p}}{4}\sin \theta _{1}\sin \theta _{2}\sin \left( \alpha
_{2}+\beta _{2}\right) +\frac{\mu _{p}}{4}\sin \theta _{1}\sin \theta
_{2}\sin \left( \alpha _{2}-\beta _{2}\right)  \notag \\
&&
\end{eqnarray}%
where%
\begin{equation*}
\mu _{p}=\frac{1+\left( 1-p\right) ^{2}}{2}
\end{equation*}%
It is obvious from above payoffs\ that quantum player Bob can always out
perform Alice, for all values of $p$. Similarly for the case of Chicken and
Battle of Sexes game, it can be proved that the classical player can be out
performed by Bob, at $\alpha _{2}=\frac{\pi }{2},\beta _{2}=0$\ and $\theta
_{2}=\frac{\pi }{2}$ and $\alpha _{2}=-\frac{\pi }{2},\beta =0$ and $\theta
_{2}=\frac{\pi }{2},$ respectively$.$

\section{Conclusion}

Quantum games with correlated noise are studied under the generalized
quantization scheme \cite{nawaz}. Three games, Prisoner Dilemma, Battle of
sexes and Chicken are studied with one player restricted to classical
strategy while other allowed to play quantum strategies. It is shown that
the effects of the memory and decoherence become effective for the case, $%
\gamma =\delta =\frac{\pi }{2},$for which quantum player out perform
classical player. It is also shown that memory controls payoffs reduction
due to decoherence and for the limit of maximum memory decoherence becomes
ineffective.

\appendix

\begin{center}
\textbf{Appendix A: Some Classical Games}
\end{center}

Here we briefly describe three classical games on interest.

\begin{center}
\textbf{Prisoner's Dilemma}
\end{center}

This game depicts a situation where two suspects (players), who have
committed a crime together, are being interrogated in a separate cell. The
two possible moves for each player are to cooperate ($C$) or to defect ($D$)
without any communication between them but having access to the following
payoff matrix.

\begin{equation}
\text{Alice}%
\begin{array}{c}
C \\ 
D%
\end{array}%
\overset{\text{Bob}}{\overset{%
\begin{array}{cc}
C\text{ \ \ \ \ } & D%
\end{array}%
}{\left[ 
\begin{array}{cc}
\left( 3,3\right) & \left( 0,5\right) \\ 
\left( 5,0\right) & \left( 1,1\right)%
\end{array}%
\right] }},  \tag{A1}  \label{matrix prisoner}
\end{equation}%
\emph{\ }It is obvious from the payoff matrix \ref{matrix prisoner} that $D$%
\ is the dominant strategy for the two players. Therefore, rational
reasoning forces the players to play $D$. Thus ($D,D$) is the Nash
equilibrium of this game with payoffs $(1,1)$. But the players could get
higher payoffs if they would have played $C$\ instead of $D$. This is the
dilemma in this game.

\begin{center}
\textbf{The Chicken game}
\end{center}

The payoff matrix for this game is

\begin{equation}
\text{Alice}%
\begin{array}{c}
C \\ 
D%
\end{array}%
\overset{\text{Bob}}{\overset{%
\begin{array}{cc}
C\text{ \ \ \ \ } & D%
\end{array}%
}{\left[ 
\begin{array}{cc}
\left( 3,3\right) & \left( 1,4\right) \\ 
\left( 4,1\right) & \left( 0,0\right)%
\end{array}%
\right] }},  \tag{A2}  \label{matrix chicken}
\end{equation}%
In this game two players drove their cars towards each other. The first one
to swerve to avoid collision is the loser (chicken) and the one who keeps on
driving straight is the winner. There is no dominant strategy in this game.
There are two Nash equilibria $\left( C,D\right) $ and $\left( D,C\right) ,$
the former is preferred by Bob and the latter is preferred by Alice. The
dilemma of this game is that the Pareto Optimal strategy $\left( C,C\right) $
is not Nash equilibrium.

\begin{center}
\textbf{Battle of Sexes}
\end{center}

The payoff matrix for this game is

\begin{equation}
\text{Alice}%
\begin{array}{c}
O \\ 
T%
\end{array}%
\overset{\text{Bob}}{\overset{%
\begin{array}{cc}
O\text{ \ \ \ \ } & T%
\end{array}%
}{\left[ 
\begin{array}{cc}
\left( 2,1\right) & \left( 0,0\right) \\ 
\left( 0,0\right) & \left( 1,2\right)%
\end{array}%
\right] }},  \tag{A3}  \label{battle of sexes}
\end{equation}%
In the usual exposition of this game Alice is fond of Opera whereas Bob
likes watching TV but they also want to spend the evening together. In the
absence of communication they face a dilemma in choosing their strategies.

\end{document}